\documentstyle[aipmod,eqalign,graphicx,times]{article}
\font\tenmib=cmmib10 \font\tensfb=cmssbx10 \relax
\def\dvp{d^3{\bf p}}\def\bbeta{\hbox{\tenmib \char'014 }}
\def\mat#1{\hbox{\tensfb #1}}
\advance \textwidth by 1in
\advance \oddsidemargin by -0.5in
\advance \textheight by 1in
\advance \topmargin by -0.5in

\makeatletter \def\@textbottom{\vskip 0pt plus 1fil minus 1pt} \makeatother

\begin{document}
\title{Efficiency of Current Drive by Fast Waves}
\author{Charles F. F. Karney and Nathaniel J. Fisch\\
Plasma Physics Laboratory, Princeton University\\Princeton, NJ 08544}
\date{PPPL--2128 (Aug.~1984)\\
Phys.\ Fluids {\bf 28}(1), 116--126 (Jan.~1985)}
\maketitle
\begin{abstract}
The Rosenbluth form for the collision operator for a weakly
relativistic plasma is derived.  The formalism adopted by Antonsen and
Chu can then be used to calculate the efficiency of current drive by
fast waves in a relativistic plasma.  Accurate numerical results and
analytic asymptotic limits for the efficiencies are given.
\end{abstract}
\section{INTRODUCTION}\label{intro}

Currents may be efficiently generated in a plasma by the injection of
rf waves whose phase velocities are several times the electron thermal
speed \cite{Fisch1}.  The efficiency, defined as the ratio of current
generated to power dissipated, is achieved in this instance because the
rf-generated plateau decays at a rate given by the collision frequency
for the fast electrons, which is relatively low.  In the quest for
higher efficiencies, current drive by waves which interact with
relativistic electrons has also been considered \cite{Fisch2}.
Relativistic effects modify the scaling of the efficiency, placing an
upper bound on the efficiency achievable by current drive by fast
waves.  In this paper, we do several things:
we give a more complete analysis of this problem
based on a formalism adopted by Antonsen and Chu \cite{Antonsen}.
Specifically, we find that the effect of finite electron temperature
leads to an enhancement of the efficiency.  In order to calculate this
effect,  we first
give expressions for the most important terms in the electron-electron
collision integral in the relativistic limit.  These expressions are put in
Rosenbluth form so as to be amenable to easy implementation on a
computer.  We imagine that the relativistic Rosenbluth potentials that
we identify may be useful in other problems arising in very hot plamas.

In order to put the present work in perspective, let us briefly review
the chief tools used in the study of current drive.  The early work used
fairly crude analytical models \cite{Fisch1,Wort}.  These models were
sufficient to obtain the scaling laws for the efficiency of current
drive, but were unable to provide the coefficients with any accuracy.
Therefore, the analytical treatment was supplemented by numerical
solutions to the two-dimensional (in momentum space) Fokker--Planck
equation \citea{\citenum{Karney-lh}--\citenum{Karney-ec}}, from which
accurate estimates of the efficiency could be found.  The first accurate
analytical treatment of current drive was based on a Langevin
formulation of the electron motion \cite{Fisch2,Fisch-Boozer}.  This
involved taking the electron temperature to be small, allowing energy
scattering to be ignored.  The moment hierarchy for the Langevin
equations can then be closed, which allows an analytical solution to be
obtained.  This was followed by a more complete numerical study of the
Fokker--Planck equation for current drive in which the problem was
reduced to the numerical solution of a one-dimensional
integro-differential equation with a source due to the rf
\cite{Cordey2}.  In this work toroidal effects were also included.
The results agreed with the Langevin analysis \cite{Fisch-Boozer} in
the limit of large phase velocities (as they should) and gave more
accurate numerical data for phase velocities comparable to or smaller
than the thermal velocity.  More recently, Antonsen and Chu
\cite{Antonsen} and, independently, Taguchi \cite{Taguchi2}, using
methods first used in the study of beam-driven currents
\cite{Hirshman,Taguchi1}, recognized that it is not necessary to solve
the rf-driven Fokker--Planck equation in order to find the rf-induced
current.  Instead, they showed that the Green's function for the current
is the Spitzer--H\"arm function \cite{Spitzer} describing the perturbed
electron distribution in the presence of an electric field.  This
reduces the problem to the determination of a single two-dimensional
function, from which the current generated by any form of rf drive can
be calculated by a simple integration.

Up until now, the only reliable analytical results for current drive in
a relativistic plasma are those obtained using the Langevin methods by
Ref.~\citenum{Fisch2}.  As we will show, these are only exact for
$T_e\ll m_ec^2$ and $p^2\gg m_eT_e$ (where $p$ is the
momentum of the resonant electrons).
A more complete analytical or numerical treatment along the
lines of that achieved in the nonrelativistic case was hampered by the
lack of a convenient form for the relativistic collision operator.  This
is remedied to some extent by the results of this paper where we
calculate the collision integrals for the first Legendre harmonic of
the perturbed electron distribution neglecting electromagnetic effects
on the binary interaction (in this approximation the collision integral
reduces to the Landau form).  Having done this, we are able to
generalize the treatment of Antonsen and Chu \cite{Antonsen} to the
relativistic case.  A number of useful results flow from this: we can
numerically calculate to high precision the current-drive efficiencies
in the relativistic regime.  We can perform an asymptotic analysis of
the Spitzer--H\"arm problem to obtain analytic approximations to the
efficiencies.  In addition, we give higher-order asymptotic corrections to
the current-drive efficiencies in the nonrelativistic limit.
Throughout this paper, toroidal effects are entirely ignored.  Although
these effects are important in the study of current drive by
low-phase-velocity waves, they play little role in current drive by
fast waves.  Incorporation of these effects, however, proceeds in exact
analogy with the treatment for the nonrelativistic case \cite{Antonsen}.

Relativistic effects on rf current drive have also been considered by
Hizanidis and Bers \cite{Hizanidis}.  They take moments of the kinetic
equation.  In order to close the resulting system of equations, they
approximate the steady-state electron distribution by a delta function.
This approximation is unjustified and, consequently, their results for
the current-drive efficiency are incorrect.

The plan of this paper is as follows: In Sec.~\ref{collision} we show
how the relativistic collision operator may be reduced to the Landau
form.  In this form, the collision operator is costly to evaluate
numerically.  So, in Sec.~\ref{rosen} we convert the collision integrals
to a Rosenbluth form, which may be evaluated very efficiently.  The
formulation of Antonsen and Chu is generalized to the relativistic case
in Sec.~\ref{formulation}.  The numerical results for the efficiencies
are given in Sec.~\ref{numerical} and the asymptotic results in
Sec.~\ref{asymptotics}.  Finally, in Sec.~\ref{high-coll}, we examine
the asymptotic form of the efficiencies using the full relativistic
collision operator.

\section{RELATIVISTIC COLLISION OPERATOR}\label{collision}

The collision operator for a relativistic plasma is given
by Beliaev and Budker \cite{Beliaev}.  They give the collision operator
as
$$\eqalignno{\left.{\partial f_a({\bf p})\over\partial t}\right|^{\rm coll}
&=\sum_b C(f_a,f_b),&(\eqlab{coll}a)\cr
C(f_a,f_b)&={q_a^2q_b^2\over 8\pi\epsilon_0^2}\log\Lambda^{a/b}
{\partial\over\partial {\bf p}}\cdot\int\mat U\cdot\biggl(
f_b({\bf p}'){\partial f_a({\bf p})\over\partial{\bf p}}-
f_a({\bf p}){\partial f_b({\bf p}')\over\partial{\bf p}'}\biggr)
\,\dvp',\quad&(\ref{coll}b)\cr}$$
where the kernel $\mat U$ is given by
$$\eqalignno{\mat U&={\gamma_a\gamma_b'(1-\bbeta_a\cdot\bbeta_b')^2\over
c[\gamma_a^2\gamma_b^{\prime2}(1-\bbeta_a\cdot\bbeta_b')^2-1]^{3/2}}
\{[\gamma_a^2\gamma_b^{\prime2}(1-\bbeta_a\cdot\bbeta_b')^2-1]\mat I\cr
&\qquad\qquad{}-\gamma_a^2\bbeta_a\bbeta_a
-\gamma_b^{\prime2}\bbeta_b'\bbeta_b'
+\gamma_a^2\gamma_b^{\prime2}(1-\bbeta_a\cdot\bbeta_b')
(\bbeta_a\bbeta_b'+\bbeta_b'\bbeta_a)\}.&(\eqlab{bb})\cr}$$
Here $a$ and $b$ are species labels, $q_s$ is the charge of species
$s$, $\log\Lambda^{a/b}$ is the Coulomb logarithm, $\epsilon_0$ is the
dielectric constant, ${\bf p}$ is the momentum, ${\bf v}_s=c\bbeta_s
={\bf p}/m_s\gamma_s$ is the velocity of species $s$,
and $\gamma_s=(1+p^2/m_s^2c^2)^{1/2}$.  The distributions are
normalized so that
$$\int f_s({\bf p})\,\dvp=n_s,$$
the number density.
We are primarily interested in situations where fast electrons are
colliding off a weakly relativistic background.  In that case
$\beta_b'\ll1$, and we can approximate $\mat U$ by its nonrelativistic
form
$$\mat U={u^2\mat I-{\bf uu}\over u^3},
\quad {\bf u}={\bf v}_a-{\bf v}_b'.\eqn(\ref{coll}c)$$
Since the original form for $\mat U$ was symmetric in the
primed and unprimed variables, we could equally well have obtained
Eq.~(\ref{coll}c) under the assumption that $\beta_a\ll1$.
The relative difference between Eqs.~(\ref{coll}c) and~(\ref{bb})
is $O(\beta_b')$.  However, the error in the collision operator
$C(f_a,f_b)$ is smaller than this.  This point is examined in more
detail in Sec.~\ref{high-coll}.
Equations~(\ref{coll}) are precisely the collision operator given by
Landau \cite{Landau}.  Indeed an examination of his derivation shows
that the mechanics of the collisions are treated relativistically; the
interaction, however, is calculated nonrelativistically assuming a
Coulomb potential.  Use of Landau collision operator implies a neglect
of the relativistic (i.e., electromagnetic) effects on the binary
interaction.  What we have shown here is that such an approximation is
valid provided at least one of the colliding particles is
nonrelativistic.

It is readily established that Eqs.~(\ref{coll})
conserve number, momentum, and
energy (${\cal E}_s=m_sc^2\gamma_s$), that an $H$-theorem applies, and
that the equilibrium solution is a relativistic Maxwellian
$f_s({\bf p})\propto\exp(-{\cal E}'_s/T)$, where
${\cal E}'_s=\penalty100 ({\cal E}_s-\penalty100 {\bf v}_d\penalty1000
\cdot\penalty1000 {\bf p})/\sqrt{1-v_d^2/c^2}$ is the energy in a frame
moving at ${\bf v}_d$, and $T$ and ${\bf v}_d$ are independent of the
species $s$.

Throughout the rest of this paper we will restrict our attention to an
electron-ion plasma.  We assume the ions are stationary and
infinitely massive ($m_i\rightarrow\infty$).  This allows us to
express the electron-ion collision operator in $(p,\mu)$ space
(where $\mu=p_\parallel/p$ and $\parallel$ and $\perp$ are with respect
to the magnetic field) as
$$C(f,f_i)=\Gamma{Z\over2vp^2}
{\partial\over\partial\mu}(1-\mu^2)
{\partial\over\partial\mu}f({\bf p}),\eqn(\eqlab{ions})$$
where
$$\eqalign{\Gamma&={n_eq_e^4\log\Lambda^{e/e}\over 4\pi\epsilon_0^2},\cr
Z&=-{q_i\log\Lambda^{e/i}\over q_e\log\Lambda^{e/e}}
\approx-{q_i\over q_e},\cr}$$
and we have assumed neutrality $q_en_e+q_in_i=0$.
In Eq.~(\ref{ions}) and henceforth we will omit the species labels
from all electron quantities.

\section{GENERALIZATION OF THE ROSENBLUTH POTENTIALS}\label{rosen}

For computational purposes, the Landau operator is not the most
convenient form for the collision operator.  If the plasma is
azimuthally symmetric, a two-dimensional integration must be performed at
each point in momentum space.  If the number of
grid points is $N\times N$, this
requires $O(N^4)$ calculations.  This requirement is dramatically
reduced in the nonrelativistic case by expressing the collision
operator in terms of Rosenbluth potentials \cite{Rosenbluth}.
Unfortunately, although the Landau operator can be used without change
to describe the collisions in a relativistic Coulomb plasma, the
Rosenbluth form no longer applies.  (The derivation of the Rosenbluth
form from the Landau form requires, for instance, that
$(\partial/\partial {\bf p}) \cdot\mat U =-(\partial/\partial {\bf p}')
\cdot\mat U$, a relation that only holds nonrelativistically.)

However, because the kernel of the collision integral Eq.~(\ref{coll}c)
has the same form as in the nonrelativistic case, it is possible to
borrow some of the techniques of Ref.~\citenum{Rosenbluth}.  We
convert the ${\bf p}'$ integration in
Eq.~(\ref{coll}b) to ${\bf v}'$ space, substitute a particular
Legendre component for $f({\bf p}')$, and manipulate the resulting
integrals into the form
$$\int \left|{\bf v}-{\bf v}'\right| P_k(\mu')h(v')\,d^3{\bf v}'$$
or
$$\int \left|{\bf v}-{\bf v}'\right|^{-1} P_k(\mu')h(v')\,d^3{\bf v}',$$
which may be evaluated in the same way as Rosenbluth potentials
\cite{Rosenbluth} ($P_k$ is a Legendre polynomial).

Here we give the resulting expressions for collisions off a
stationary Maxwellian background, i.e., $C(f,f_m)$,
and for collisions of a Maxwellian off the first Legendre
component of a background, i.e., $C(f_m,\mu f_1)$.  In both cases only
electron-electron collisions are considered.  These terms are all that
are required for the solution of the Spitzer--H\"arm problem (giving the
Green's function for the rf current drive) and they suffice for an
accurate numerical solution of the two-dimensional Fokker--Planck
equation as described in Sec.~\ref{numerical}.

Beginning with the case of collisions off a Maxwellian, let us start
by assuming merely that the background is isotropic $f({\bf p})=f_0(p)$.
The three-dimensional
integrals in Eq.~(\ref{coll}b) then reduce to one-dimensional integrals
giving
$$C(f,f_0)={1\over p^2}{\partial\over\partial p}p^2
\Bigl(A(p){\partial\over\partial p}+F(p)\Bigr)f({\bf p})
+{B(p)\over p^2}{\partial\over\partial\mu}(1-\mu^2)
{\partial\over\partial\mu}f({\bf p})\eqn(\eqlab{isotrop}a)$$
where
$$\eqalignno{
A(p)&={4\pi \Gamma\over 3n}
\biggl[\int_0^p p^{\prime2}f_0(p') {v^{\prime2}\over v^3}\,dp'+
\int_p^\infty p^{\prime2}f_0(p') {1\over v'}\,dp'\biggr],
&(\ref{isotrop}b)\cr
F(p)&={4\pi \Gamma\over 3n}
\biggl[\int_0^p p'f_0(p') {3v'-v^{\prime3}/c^2\over v^2}
\,dp'+
\int_p^\infty p'f_0(p') 2v/c^2\,dp'\biggr],\qquad
&(\ref{isotrop}c)\cr
B(p)&={4\pi \Gamma\over 3n}
\biggl[\int_0^p p^{\prime2}f_0(p') {3v^2-v^{\prime2}\over 2v^3}
\,dp'+
\int_p^\infty p^{\prime2}f_0(p') {1\over v'}\,dp'\biggr].
&(\ref{isotrop}d)\cr
}$$
Specializing to the case $f_0=f_m$ and using the relation
$\partial f_m/\partial p=-(v/T)f_m$, we find that
$$F(p)=(v/T)A(p)$$ and the steady-state solution to $C(f,f_m)=0$ is
that $f$ is a relativistic Maxwellian \cite{DeGroot} with temperature $T$
$$f_m(p)={n\over4\pi m^2cTK_2(\Theta^{-1})}
\exp(-{\cal E}/T),\eqn(\eqlab{maxwell})$$
where
$$\eqalign{{\cal E}&=mc^2\gamma,\cr
\Theta&= {T\over mc^2}\cr}$$
($\Theta=1$ corresponds to an electron temperature of $511\,\hbox{keV}$),
and $K_n$ is the $n$th-order modified Bessel function of the second
kind.

For later use we define here a thermal momentum
$$p_t=\sqrt{mT},$$
a mean-squared velocity
$$\eqalign{v_t^2&={1\over 3n}\int v^2
f_m(p)\,\dvp={T\over m}V_t^2,\cr
V_t^2&=1-{5\over2}\Theta
+{55\over8}\Theta^2+O(\Theta^3),\cr}$$
a thermal collision frequency
$$\nu_t={m\Gamma\over p_t^3}=
{n q^4m\log\Lambda\over 4\pi\epsilon_0^2 p_t^3},$$
and a collision frequency normalized to the speed of light
$$\nu_c={\Gamma\over m^2 c^3}=
{n q^4\log\Lambda\over 4\pi\epsilon_0^2 m^2 c^3}.$$
These frequencies differ by a factor of two from those used
in earlier publications
\cite{Fisch1,Fisch2,Karney-lh,Fisch-Karney,Karney-ec,Fisch-Boozer}.
Specifically, we have
$\nu_t=\nu_0/2$ and $\nu_c=\nu/2$.  This means that all our normalized
efficiencies  are a factor of two smaller than in these earlier papers.
(We made this change because the normalized Fokker--Planck equation in
the high-energy limit now has a simpler form.  This convention is also
used by other workers in this field.)

For $p\gg p_t$, the indefinite limits in the integrals
in Eq.~(\ref{isotrop}) can be replaced by $\infty$, giving
\cite{Mosher}
$$\eqalignno{A(p)&=\Gamma {v_t^2\over v^3},&(\eqlab{high}a)\cr
B(p)&=\Gamma{1\over 2v}\biggl(1-{v_t^2\over v^2}\biggr).
&(\ref{high}b)\cr}$$
Note that the frictional force $F(p)$ reaches a constant value as
$p\rightarrow\infty$.  This implies, for instance, that an electric
field smaller than $\Gamma v_t^2/qTc^2$ cannot produce runaways
\cite{Connor}.  On the other hand, the pitch-angle scattering frequency
$B(p)/p^2$ continues to decay as $p\rightarrow\infty$.  As the energy of
the electron increases, its effective mass increases. It is then more
difficult to deflect the heavier particle.  In this limit, pitch-angle
scattering is negligible compared with frictional slowing down.  This is
to be contrasted with the nonrelativistic case where the pitch-angle
scattering frequency and the frictional slowing-down rate decay as $1/p^2$
and the two processes are of comparable importance.

The implication for current drive is that the efficiency of parallel
wave-induced fluxes, say by lower-hybrid waves, approaches a constant.
This can be seen as follows: Nonrelativistically, the efficiency
increases as $p^2$.  Relativistic electrons, however, slow down faster
because they are heavier, and they also do not carry more current when
pushed in the parallel direction.  Each of these effects reduces the
efficiency by $\gamma\sim p$; hence the approach to a constant.

The other term we shall need is $C(f_m,\mu f_1)$.  This term is rather
harder to compute.  We define
$f_1(p)=f_m(p)\chi_1(p)$ and write $C(f_m,\mu f_m\chi_1)=
\mu f_m I(\chi_1)$.  Again, we reduce (this time after much algebra)
the integrals
in Eq.~(\ref{coll}b) to one-dimensional ones to give
$$\eqalignno{I(\chi_1)={4\pi \Gamma\over n}
&\Biggl\{{m f_m(p)\chi_1(p)\over \gamma}\cr
&\quad{}+{1\over 5}\int_0^p p^{\prime2}f_m(p')\chi_1(p'){m\over T}\biggl[
{\gamma\over p^2}{v'\over\gamma^{\prime3}}
\biggl({T\over mc^2}(4\gamma^{\prime2}+6)
-{1\over3}(4\gamma^{\prime3}-9\gamma')\!\biggl)\cr
&\qquad\qquad\qquad\qquad\qquad\qquad\quad
{}+{\gamma^2\over p^2}{v'\over\gamma^{\prime3}}
\biggl({mv^{\prime2}\over T}\gamma^{\prime3}
-{1\over3}(4\gamma^{\prime2}+6)\!\biggl)
\biggr]\,dp'\cr
&\quad{}+{1\over 5}\int_p^\infty p^{\prime2}f_m(p')\chi_1(p'){m\over T}\biggl[
{\gamma'\over p^{\prime2}}{v\over\gamma^{3}}
\biggl({T\over mc^2}(4\gamma^{2}+6)
-{1\over3}(4\gamma^{3}-9\gamma)\!\biggl)\cr
&\qquad\qquad\qquad\qquad\qquad\qquad\quad
{}+{\gamma^{\prime2}\over p^{\prime2}}{v\over\gamma^{3}}
\biggl({mv^{2}\over T}\gamma^{3}
-{1\over3}(4\gamma^{2}+6)\!\biggl)
\biggr]\,dp'\Biggr\}.&(\eqlab{legend1})\cr}$$
The term in square brackets in the last integral matches that in the
first integral except for the interchange of the primed and unprimed
variables.  The simplification of Eq.~(\ref{legend1}) was achieved, in
part, with the help of the symbolic manipulation program, {\small
MACSYMA} \cite{MACSYMA}.

Equations~(\ref{isotrop}) and (\ref{legend1}) are now in a
computationally convenient form.  Their evaluation involves the
determination of a number of indefinite integrals (the unprimed
variables should be factored out of the integrals for this step), and
the multiplication of these integrals by various functions of $p$.  If
the distribution functions are known on a grid of $N$ points, then the
computational cost is just $O(N)$.  Furthermore, the calculation can be
arranged so that nearly all the computations vectorize \cite{McCoy}.
The general solution of the linearized electron-electron collision
operator $C(f,f_m) + C(f_m,f)=0$ is
$$f=(a+{\bf b}\cdot{\bf p}+c{\cal E})f_m,$$
where $a$, $\bf b$, and $c$ are arbitrary constants.  With $a=c=0$ and
${\bf b}=\hat{\bf p}_\parallel$, this provides a
useful check on Eqs.~(\ref{isotrop}) and (\ref{legend1}) and their
computational realizations.

\section{FORMULATION}\label{formulation}

We now turn to the calculation of the rf efficiency.  There are three
steps involved: the specification of the rf current-drive problem, the
identification of the Spitzer--H\"arm function as the Green's function
for the rf-driven current; and the solution of the Spitzer--H\"arm
problem.

We begin with the specification of the problem.  This is just a standard
application of the Chapman--Enskog procedure \cite{Chapman}.  The most
important assumption is that the collisional time scale is much shorter
than the transport time scale (the time scale for heating the plasma by
the rf).  This places some restrictions on the rf drive.  However, these
are usually not severe ones in the case of fast-wave current drive
because, even if the rf is strong, there are few resonant particles
and, consequently, the heating rate is small.

The
effect of the rf is to induce an electron flux
$${\bf S}=-\mat D\cdot
{\partial f\over\partial\bf p}\eqn(\eqlab{flux})$$
in momentum space, where $\mat D$ is the quasilinear diffusion tensor
\cite{Kennel}.  In the Chapman--Enskog ordering this is taken to be of
first order.  The zeroth-order electron distribution is given by setting
the collision term $C(f,f)+C(f,f_i)$ equal to zero.  The general
solution is a Maxwellian Eq.~(\ref{maxwell}) with $n$ and $T$ arbitrary
functions of time and position.  For simplicity we ignore the spatial
variations.  Since the rf drive is particle conserving, we may take $n$
to be a constant.  A drifting Maxwellian does not solve the zeroth-order
system since the ions are taken to be stationary.

The first-order equation is given by substituting $f=f_m(1+\psi)$ with
$\psi$ ordered small to give
$$C(f_m\psi)=
{\partial\over\partial \bf p}\cdot{\bf S}
+{({\cal E}-\left<{\cal E}\right>)\over T}
f_m{d\over dt}\log T,
\eqn(\eqlab{rf})$$
where
$$C(f)=C(f,f_m)+C(f_m,f)+C(f,f_i)\eqn(\eqlab{lin-coll})$$
is the linearized collision operator, and
$\left<{\cal E}\right>$ is the mean energy per particle \cite{DeGroot}
$$\eqalign{\left<{\cal E}\right>
&={1\over n}\int {\cal E}f_m(p)\,\dvp\cr
&=mc^2\biggl({K_1(\Theta^{-1})\over K_2(\Theta^{-1})}
+3\Theta\biggr).\cr}$$
The last term in Eq.~(\ref{rf})
represents the heating of the Maxwellian.  The equation
for the time evolution of $T$ is given by the solubility condition
for Eq.~(\ref{rf}), which is obtained by taking its energy moment.
Since
the linearized collision operator is energy conserving (recall that
we take
the limit $m_i\rightarrow\infty$, so that there is no energy
exchange between electron and ions), this gives
$$n{d\left<{\cal E}\right>\over dt}=P$$
where $P$ is the power dissipated per unit volume by the rf
$$P=\int {\bf S}\cdot{\bf v}\,\dvp.\eqn(\eqlab{p-d})$$
[There is another solubility condition given by the density moment
of Eq.~(\ref{rf}).  This is automatically satisfied by taking $dn/dt=0$.]
The solution to Eq.~(\ref{rf}) is made unique by demanding that
$f_m\psi$ have zero density and energy.

In the nonrelativistic limit, Eq.~(\ref{rf}) is the equation solved
numerically by Cordey {\it et al} \cite{Cordey2}.
However, since we are usually interested primarily in the current
density generated by the rf
$$J=q\int v_\parallel f_m\psi\,\dvp,\eqn(\eqlab{current-def})$$
and the efficiency of current generation defined by the ratio
$J/P$,
we usually do not need to know the full solution for $\psi$.

The method for determining the current without solving for $\psi$ was
given by Hirshman \cite{Hirshman} and by Taguchi \cite{Taguchi1} for
neutral-beam-driven currents and was introduced into the study of
rf-driven currents by Antonsen and Chu \cite{Antonsen} and Taguchi
\cite{Taguchi2}.  The key is to define an ``adjoint'' problem
$$C(f_m\chi)=-q v_\parallel f_m.\eqn(\eqlab{spitz})$$
Again $f_m\chi$ is required to have zero density and energy.
This is the Spitzer--H\"arm problem for the perturbed electron
distribution function due to an electric field
${\bf E}=T\hat{\bf p}_\parallel$.
Using the self-adjoint property of the linearized collision operator
$\int\psi C(f_m\chi)\,\dvp=\int\chi C(f_m\psi)\,\dvp$, it is readily
found that
$$J=\int{\bf S}\cdot{\partial\over\partial \bf p}
\chi({\bf p})\,\dvp.\eqn(\eqlab{current})$$
In this equation $\chi$ plays the role of a Green's function for the
rf-driven current.
The ratio of Eqs.~(\ref{current}) and (\ref{p-d}) gives the efficiency
$${J\over P}={{\displaystyle
\int{\bf S}\cdot{\partial\over\partial \bf p}\chi({\bf p})
\,\dvp}\over{\displaystyle
\int {\bf S}\cdot{\bf v}\,\dvp}}.\eqn(\eqlab{effa})$$
An important special case is when the rf excitation is localized.  Then
it is only necessary to know the position and direction of the
excitation to determine the efficiency
$${J\over P}={{\displaystyle
\hat{\bf S}\cdot{\partial\over\partial \bf p}\chi({\bf p})}
\over{\displaystyle
\hat{\bf S}\cdot{\bf v}}},\eqn(\eqlab{effb})$$
where all quantities are now evaluated at the position of the
excitation.  If we compare this method with the Langevin method of
Fisch \cite{Fisch2}, we see that $\chi$ is the mean-integrated current
due to a group of electrons released at $\bf p$ at $t=0$
$$\chi({\bf p})=q\int_0^\infty \left<v_\parallel\right>\,dt.$$

The power of these results is that the calculation of $J/P$ 
does not require a solution of
Eq.~(\ref{rf}) for the rf distribution $\psi$.  On the other hand,
Eq.~(\ref{spitz}) must be solved for the
Spitzer--H\"arm function $\chi$.  This reduces to the solution of a
one-dimensional integro-differential equation, which may be
accurately computed.  Furthermore, in the nonrelativistic limit, it
has been tabulated \cite{Spitzer}.  This method also substantially
reduces the parameter space to be investigated numerically.  The
solution of Eq.~(\ref{spitz}) depends on two parameters only, $Z$ and
$\Theta$.  In contrast, the solution of Eq.~(\ref{rf}) depends on
various parameters specifying the nature of the rf excitation (for
instance, the direction of $\bf S$, the minimum and maximum phase
velocities, etc.)\ as well as $Z$ and $\Theta$.

In order to determine the rf current-drive efficiency using
Eqs.~(\ref{effa}) or~(\ref{effb}), we must solve the Spitzer--H\"arm
problem, Eq.~(\ref{spitz}).  The solution $\chi$ consists of only the
first Legendre harmonic, so we substitute $\chi({\bf p})=\mu\chi_1(p)$ into
Eq.~(\ref{spitz}) giving
$${1\over p^2}{\partial\over\partial p}p^2A(p)
{\partial\chi_1\over\partial p}
-{vA(p)\over T}{\partial\chi_1\over\partial p}
-{2B(p)+\Gamma Z/v\over p^2}\chi_1+I(\chi_1)+qv=0,\eqn(\eqlab{spitz1})$$
where $A(p)$ and $B(p)$ are given by Eq.~(\ref{isotrop}), the
electron-ion term is given by Eq.~(\ref{ions}), and $I(\chi_1)$ is
given by Eq.~(\ref{legend1}).  The fact that the solution of
Eq.~(\ref{spitz}) consists of only a single Legendre component
constitutes an additional advantage to this method of determining current-drive
efficiencies.  The solution of the full rf problem given in
Eq.~(\ref{rf}) consists, in general, of many Legendre components.  Often
some truncation is performed in computing these numerically.

Equation~(\ref{spitz1}) may be solved by approximate analytic methods
either by expressing $\chi$ as a sum of Sonine polynomials
\cite{Chapman,Braginskii} or by formulating the equation as a
variational problem \cite{Hirshman}.  These methods have the
disadvantage that they generally fail to reproduce the correct
asymptotic (large $\bf p$) form for $\chi$.  This failing does not
affect the calculation of the electrical conductivity significantly
since in that case $\chi$ is integrated with a weighting factor
proportional to $f_m$.  However, it rules out such methods for the study
of rf current drive, since the efficiency may depend on the local value
of $\chi$.

This leaves us either with asymptotic methods, which we apply in
Sec.~\ref{asymptotics}, or with numerical methods.  Numerical solutions
to Eq.~(\ref{spitz1}) have been given in the nonrelativistic case in
Refs.~\citenum{Spitzer} and \citenum{Cordey1}.  Here we use a simpler
method that avoids most of the problems with the application of
boundary conditions.  We cast Eq.~(\ref{spitz1}) as a one-dimensional
diffusion equation by setting the left-hand side to $\partial \chi_1/
\partial t$ and solve this diffusion equation until a steady state is
reached.  (The initial conditions may be chosen arbitrarily.)  The
integration is carried out in the domain $0<p<p_{\rm max}$ and the boundary
conditions $\chi(0)=0$ and $\chi''(p_{\rm max})=0$ are imposed.  The
diffusion equation describes the physical problem of the evolution of
the perturbed electron distribution in the presence of an electric
field and is therefore guaranteed to give the correct solution of
Eq.~(\ref{spitz}) without having to worry about spurious solutions
that diverge at $p=0$ or $p=\infty$.  Since this is a one-dimensional
diffusion equation, it may be solved by treating the differential
operator fully implicitly (the time step may be taken to be large).  The
integral operator $I(\chi_1)$ is treated explicitly and is recomputed
after every time step.  In the calculations shown here, the momentum
step size was taken to be $p_t/50$, the time step was taken to be
$1000/\nu_t$, and the process converged (i.e., the relative change in
$\chi_1$ per step was less than 1 part in $10^{10}$) after about $50$
steps.

In the following sections we will also need the function
$G(p)=\chi_1(p)/p$ so that $\chi({\bf p})=p_\parallel G(p)$.  In terms
of $G$, the gradient of $\chi$ is
$${\partial\over\partial \bf p}\chi({\bf p})=G(p)\hat{\bf p}_\parallel+
p_\parallel G_p(p) \hat{\bf p},$$
where $G_p(p)=dG(p)/dp$.

\section{NUMERICAL RESULTS}\label{numerical}

The solution for $\chi$ is given as a contour plot in
Fig.~\ref{contour} for $Z=1$ and $\Theta=0$ and 0.01.  From these and a
knowledge of $\bf S$, the direction of the rf-induced current can be
determined.  In the nonrelativistic case, Fig.~\ref{contour}(a), $\chi$
rises ever more steeply as $p$ is increased, giving the favorable $p^2$
scaling for the current-drive efficiency \cite{Fisch-Boozer}.  On the
other hand, in a hot plasma, Fig.~\ref{contour}(b), the slope reaches a
constant (the contour levels are equally spaced), leading to a limit in
the efficiency of the current drive \cite{Fisch2}.

Figure~\ref{contour} also shows that the contours become vertical for
$p_\parallel$ small.  This indicates that pushing electrons with small
$p_\parallel$ in the perpendicular direction (as with cyclotron-damped
waves) is not effective in generating current.  Pushing electrons in the
parallel direction is effective, especially for small $p_\parallel$,
since the denominator in Eq.~(\ref{effb}) can be small.  In general,
when the contours of constant energy ($p=\hbox{constant}$) cross
contours of constant $\chi$, the efficiency can be very large.

Turning now to the numerical results for the efficiency, we begin with
the case of a localized spectrum, Eq.~(\ref{effb}).  Although this
situation may not be realized in practice, it is important because it
can help us to determine the best current-drive schemes by showing
where in velocity space to induce the flux.  There are two major
classes of fast waves that have been considered for current drive,
namely Landau-damped waves (e.g., lower-hybrid waves)
for which $\hat{\bf S}=\hat{\bf p}_\parallel$
and cyclotron-damped waves for which  $\hat{\bf S}=\hat{\bf p}_\perp$.
Taking the limit $p_\perp\rightarrow0$, we have
$$\eqalignno{{J\over P}&={G(p)+p G_p(p)\over v}&(\eqlab{effloc}a)\cr
{J\over P}&={p G_p(p)\over v}&(\ref{effloc}b)\cr}$$
for Landau-damped and cyclotron-damped waves, respectively.  The
efficiencies are plotted in Fig.~\ref{local} for $Z=1$ and $\Theta=0$,
0.02, 0.05, 0.1, and 0.2 (these correspond to $T=0$, 10, 26,
51, and $102\,\hbox{keV}$).  The curves for $\Theta=0$ in the two
cases are given analytically from Eq.~(\ref{anal}); they agree
with the results of Ref.~\citenum{Fisch2}.  This confirms the earlier
analysis and shows that it is exact in the limit of $T\ll mc^2$ and
$p^2\gg mT$.

We next consider current drive by a narrow spectrum of Landau-damped
waves.  In this case, all particles satisfying the Landau resonance
condition $\omega-k_\parallel v_\parallel=0$ interact with the wave, and
the quasilinear diffusion tensor is
$$\eqalign{\mat D&\propto\delta(\omega-k_\parallel v_\parallel)
\hat{\bf p}_\parallel\hat{\bf p}_\parallel\cr
&\propto\gamma\,\delta(p_\parallel-mv_p\gamma)
\hat{\bf p}_\parallel\hat{\bf p}_\parallel,\cr}$$
where $v_p=\omega/k_\parallel$
is the parallel wave phase velocity.  Assuming that the electron
distribution is weakly perturbed, we can take $f=f_m$ in
Eq.~(\ref{flux}) to give
$${\bf S}\propto\gamma\,v_\parallel f_m\,
\delta(p_\parallel-mv_p\gamma)\hat{\bf p}_\parallel.$$
When we substitute this expression into Eq.~(\ref{effa}), we obtain
$${J\over P}={1\over v_p}\,{{\displaystyle \int_{p_0}^\infty\Bigl(G(p)+{(m\gamma
v_p)^2\over p}G_p(p)\Bigr)\gamma f_m(p) p\,dp}\over{\displaystyle
\int_{p_0}^\infty\gamma f_m(p) p\,dp}},\eqn(\eqlab{effnarrow})$$
where
$p_0= mv_p/(1-v_p^2/c^2)^{1/2}$ is the minimum resonant momentum.
This efficiency is plotted in Fig.~\ref{narrow}.
In the limit $v_p\rightarrow0$, the efficiency
becomes large.  This demonstrates that current may be efficiently driven
by low phase velocity waves as was proposed by Wort \cite{Wort}.

A similar analysis can be performed for a narrow spectrum of
cyclotron-damped waves.  The situation is more complicated here
because the electron cyclotron frequency depends relativistically on
the momentum \cite{Cairns} and because relativistic effects distort the
diffusion paths \cite{Karney-ec}.  In addition, the variation of the
diffusion coefficient with $p_\perp$ depends on the harmonic number.
This means that the efficiency depends on three wave parameters
$\omega/k_\parallel$, $\Omega/k_\parallel$ ($\Omega$ is the rest-mass
cyclotron frequency), and the harmonic number.  We therefore will only
treat this case in the nonrelativistic limit.

In the nonrelativistic limit ($\Theta\rightarrow0$, $p/mc\rightarrow0$),
the efficiencies for both kinds of waves have been calculated by
Cordey {\it et al.} \cite{Cordey2} and Taguchi \cite{Taguchi2}.  They
considered a narrow spectrum of Landau-damped waves for which the
efficiency is given by the nonrelativistic limit of
Eq.~(\ref{effnarrow}) and a narrow spectrum of
cyclotron-damped waves for which the diffusion coefficient is
$$\mat D\propto v_\perp^{2(l-1)}\delta(v_\parallel-v_p)
\hat{\bf p}_\perp\hat{\bf p}_\perp,$$
where $l$ is the harmonic number and
$v_p=(\omega-l\Omega)/k_\parallel$.  Assuming that $f=f_m$ in
Eq.~(\ref{flux}), the efficiency for cyclotron-damped waves is
$${J\over P}=m^2v_p{{\displaystyle \int_{p_0}^\infty
(p^2-p_0^2)^lf_m(p)G_p(p) \,dp}\over
{\displaystyle \int_{p_0}^\infty (p^2-p_0^2)^lf_m(p)p \,dp}},
\eqn(\eqlab{effec})$$
where $p_0=mv_p$.  (Here we consider only the
fundamental cyclotron resonance $l=1$.)
In Fig.~\ref{nonrel-fig}, we plot these efficiencies normalized
to the thermal quantities together with the
asymptotic results, Eqs.~(\ref{narrow-nonr}) and~(\ref{narrow-ec}a).
For $mv_p\gg
p_t$, the efficiencies scale as $v_p^2$ as predicted by Fisch and Boozer
\cite{Fisch-Boozer}.  The $1/v_p$ scaling seen in the Landau-damping case
for $mv_p\ll p_t$ is obtained by taking the limit
$v_p\rightarrow0$ in Eq.~(\ref{effnarrow}) to give
$${J\over P}={1\over v_p}\,
{\int D(p_\perp)f_m(p_\perp)G(p_\perp)p_\perp\,dp_\perp\over
\int D(p_\perp)f_m(p_\perp)p_\perp\,dp_\perp}.\eqn(\eqlab{low-freq-eff})$$
Here we have included an arbitrary dependence of $\mat D$ on $p_\perp$.
In Ref.~\citenum{Fisch-Karney}, three different types of
low-phase-velocity current drive were identified, namely by
Landau damping, transit-time magnetic pumping, and Alfv\'en waves.
These methods differ in the forms for $D(p_\perp)$
$$D(p_\perp)=\cases{
1                    &(Landau damping),\cr
(p_\perp/p_t)^4      &(transit-time magnetic pumping),\cr
[2-(p_\perp/p_t)^2]^2&(Alfv\'en waves).\cr}$$
The case plotted in Fig.~\ref{nonrel-fig} is the first one (Landau damping).
Evaluating the integrals in these cases gives
$${J\over P}=\left\{\matrix{C_L\cr C_M\cr C_A\cr}\right\}
{q\over mv_p\nu_t},$$
where the coefficients $C$ are given in Table~\ref{low-freq}.
The coefficients for $Z=1$ should be compared
with the (less exact) results of
Ref.~\citenum{Fisch-Karney} obtained by a numerical solution of the
two-dimensional Fokker--Planck equation where the constants of
proportionality are given as 4, 6.5, and 6.5, respectively.  The
coefficient $C_L$ has been determined analytically by Cordey {\it et
al.} \cite{Cordey2} to be
$$C_L={3\sqrt{2\pi}\over 2 Z}.$$
The dependence on $Z$ indicates that the current is unaffected by
electron-electron collisions.
This result may be derived by taking the momentum moment of
Eq.~(\ref{spitz}).  The electron-electron collision term then drops out
(from momentum conservation) and the electron-ion term is proportional
to the numerator in Eq.~(\ref{low-freq-eff}).

The last numerical example is one in which we relax the condition that
$f=f_m$ in Eq.~(\ref{flux}).  This allows us to find the flux $\bf S$
that develops in the presence of high rf power.
In order to determine $\bf S$, we
numerically solve the two-dimensional Fokker--Planck equation
$${\partial f\over\partial t}=C_{\rm num}(f)+
{\partial\over\partial\bf p}\cdot \mat D\cdot
{\partial f\over\partial\bf p},\eqn(\eqlab{2D})$$
until a steady state is reached.
The numerical collision operator is defined as
$$C_{\rm num}(f)=C(f,f_m)+C(f_m,\mu f_1)+C(f,f_i),$$
where $\mu f_1$ is the first Legendre harmonic of $f$.
The electron-ion term $C(f,f_i)$ is calculated using Eq.~(\ref{ions}).

In order to justify our handling of the
electron-electron collisions, let us consider the linearized
electron-electron operator $C(f,f_m)+C(f_m,f)$.
The first term describes the relaxation of the tail particles on the
bulk and the second describes the concomitant heating of the bulk.
The linearization is justified even with strong rf, as long as $f({\bf p})
\approx f_m(p)$ for ${\cal E}\sim T$.  The linearized electron-electron
operator conserves energy, and if this were used in Eq.~(\ref{2D}),
there would be nothing to balance the power input by the rf (there is
no transfer of energy to the ions in the limit $m_i\rightarrow\infty$),
and so a steady-state solution to Eq.~(\ref{2D}) would not be
possible. In Eq.~(\ref{rf}), this is handled by allowing the
temperature of the Maxwellian to increase slowly with time.  In the
numerical code, we adopt a different approach, namely to modify the
collision operator so that energy is lost in an innocuous way.  The term
responsible for the bulk heating is the second term $C(f_m,f)$.  Let
us write $f$ in this term as a Legendre harmonic expansion
$$f({\bf p})=\sum_{k=0}^\infty P_k(\mu)f_k(p).$$
Of the terms in this series, only one, the $k=0$ term, contributes
to the bulk heating.  (The energy moments of the other terms vanish.)
Thus in order to lose energy we drop the term $C(f_m,f_0)$.  Of
the remaining terms in the series, only the first, the $m=1$ term,
is of importance---it is responsible for ensuring conservation of
momentum.  Thus we retain only this term and approximate
$C(f_m,f)$ by $C(f_m,\mu f_1)$ to give the collision
operator $C_{\rm num}$.

The collision operator $C_{\rm num}$ has the following properties:
energy is not conserved (thus allowing a steady state to be reached);
momentum is conserved; and quantities such as the
Spitzer-H\"arm conductivity, which are given solely in terms of the
first Legendre harmonic, are correctly given.  To justify the way in
which energy conservation is handled, we may check that the results are
insensitive to the details of how this is done.  One such check is given
below where we compare the efficiency given by the numerical solution
of Eq.~(\ref{2D}), in which energy is lost, and that given by
Eq.~(\ref{effa}), where energy is conserved.

We assume that the rf diffusion term in Eq.~(\ref{2D}) is caused by
high-power lower-hybrid waves whose phase velocities lie between $v_1$
and $v_2$.  Thus we take
$$\mat D=\cases{D({\bf p})\hat{\bf p}_\parallel\hat{\bf p}_\parallel,&
for $v_1 < p_\parallel/(m\gamma) < v_2$,\cr
0,&otherwise\cr}$$
where $D({\bf p})$ is chosen to be large enough to plateau $f$.
[Here we choose $D({\bf p})=10\,\nu_t p_t^2 /(1+p/p_t)$.]

Figure~\ref{2D plot} shows the steady-state
solution of Eq.~(\ref{2D}) for $Z=1$,
$\Theta=0.01$ ($T\approx 5\,\hbox{keV}$), $v_1=0.4c=4p_t/m$, and
$v_2=0.7c=7p_t/m$ (the parallel refractive index satisfies
$1.43<n_\parallel<2.5$).  Using the numerical solution for $f({\bf p})$
and ${\bf S}({\bf p})$, and the definitions~(\ref{p-d})
and~(\ref{current-def}), we obtain $J= 3.74\times10^{-4}qnc$,
$P=1.28\times10^{-3}mnc^2\nu_c$, and ${J/P}=0.293\,q/mc\nu_c$.

This is to be compared with the result given by Eq.~(\ref{current})
with the numerically determined flux ${\bf S}({\bf p})$, namely $J
=3.77\times10^{-4}qnc$ and ${J/P}=0.296\,q/mc\nu_c$.  (The figure for
$P$ remains unchanged since this depends on $\bf S$ alone.)  These two
sets of figures are within 1\% of each other.  The excellent agreement
illustrates two points:  the approximations made in the numerical collision
operator, namely, the neglect of the heating term $C(f_m,f_0)$, has
little effect on the results for the current-drive efficiencies
(discretization effects are probably a greater source of error
in these results);  and the analytic result
Eq.~(\ref{effa}) can be used to obtain reliable figures for
the efficiency for cases of strong rf.  What is needed in the latter
instance is an estimate for the rf flux $\bf S$.  This may be found
from a numerical solution of a two-dimensional Fokker--Planck equation
(as here) or from an approximate analytical solution.  Some saving may
be possible using this method in conjunction with a numerical code:
since $\bf S$ reaches a steady state sooner than $f$, it may not be
necessary to run the code so long in order to obtain a reasonably
accurate estimate for the efficiency.

\section{ASYMPTOTIC ANALYSIS}\label{asymptotics}

We have seen that the efficiency of current drive may be expressed in
terms of the solution of the Spitzer--H\"arm problem,
Eq.~(\ref{spitz1}).  This equation may be approximately solved in the
limit $p\gg p_t$.  We will begin with the relativistic case and later
treat the nonrelativistic limit.  We start by writing down the
normalized form of Eq.~(\ref{spitz1}) in the limit $p\gg p_t$.  We
chose normalizations based upon $q$, $m$, $c$, and $\nu_c$.
Thus momenta are normalized to $mc$, $\chi_1$ to $qc/\nu_c$,
$J/P$ to $q/mc\nu_c$, etc.  We use the same symbols to
represent the normalized and unnormalized quantities.  The coefficients
$A(p)$ and $B(p)$ are given by Eqs.~(\ref{high}),
suitably normalized. The integral term  may be evaluated by
replacing the indefinite limits in Eq.~(\ref{legend1}) by $\infty$,
giving when normalized
$$I(\chi_1)\approx\Theta^{3/2}\biggl({H_a(\Theta,Z)\over vp}
+{H_b(\Theta,Z)\over v^2}\biggr)$$
where $H_a$ and $H_b$ are definite integrals of $\chi_1$ (and thus
independent of momentum) that must be determined
numerically.  In the limit $\Theta\rightarrow0$, both $H_a$ and $H_b$
are finite.  In normalized
form with $p^2\gg \Theta$, Eq.~(\ref{spitz1}) reads
$$\displaylines{\quad
{\Theta V_t^2\over v^3}\biggl[ {\partial^2\chi_1\over \partial p^2}-
\biggr({v\over \Theta}+{3\over v\gamma^3}-{2\over p}\biggr)
{\partial\chi_1\over \partial p}\biggr]\hfill\cr
\hfill{}-{1\over vp^2}\biggl(1+Z-{\Theta V_t^2\over v^2}\biggr)\chi_1
+\Theta^{3/2}\biggl({H_a\over vp}+{H_b\over v^2}\biggr)+v=0.
\quad(\eqlab{spitz-norm})}$$
The error in this equation is exponentially small.

We now make a subsidiary expansion in small $\Theta$.  In the limit
$\Theta\rightarrow0$, several terms in Eq.~(\ref{spitz-norm}) drop out
leaving
$$-{1\over v^2}{\partial\chi_1\over \partial p}-
{1+Z\over vp^2}\chi_1+v=0.$$
This may by solved with the boundary condition $\chi_1(p=0)=0$ to give
$$\chi_1=\biggl({\gamma+1\over\gamma-1}\biggr)^{\textstyle\!\!{1+Z\over2}}
\int_0^p\biggl({\gamma'-1\over\gamma'+1}\biggr)^{\textstyle\!\!{1+Z\over2}}
v^{\prime3}\,dp'.\eqn(\eqlab{anal})$$
This is the result derived using the Langevin equations
by Fisch \cite{Fisch2}.  For
integer values of $Z$, the integral may be expressed in terms of
elementary functions.  In particular for $Z=1$ we have
$$\chi_1=\biggl({\gamma+1\over\gamma-1}\biggr)(vp-2\log\gamma).$$
Of particular interest is the efficiency for large $p$
since this gives a limit to the efficiency of
current drive by fast waves.  If we let $p\gg1$, the integral may be
approximately evaluated to give
$$\chi_1\rightarrow p-(1+Z)\log p.$$

If we now take $\Theta$ to be finite, Eq.~(\ref{spitz-norm})
cannot be easily solved.  However, we may solve it in the limit
$p\gg1$. We achieve this by writing
$$\chi_1\approx\alpha p+\beta\log p\eqn(\eqlab{chiasym})$$
in analogy to the situation with $\Theta=0$.  Substituting this form of
$\chi_1$ into Eq.~(\ref{spitz-norm}) and balancing terms of equal order
in $p$ gives
$$\alpha={1+\Theta^{3/2}H_b\over V_t^2}\eqn(\eqlab{coeff}a)$$
from the $O(p^0)$ terms and
$$\beta=-{(1+Z-3\Theta V_t^2)\alpha-\Theta^{3/2}H_a\over V_t^2}
\eqn(\ref{coeff}b)$$
from the $O(p^{-1})$ terms.  When the rf excitation is localized, the
current-drive efficiency is given by Eqs.~(\ref{effloc}) that, with
$\chi_1$ given by Eq.~(\ref{chiasym}), read
$$\eqalignno{{J\over P}&\approx\alpha+{\beta\over p}&(\eqlab{effasym}a)\cr
{J\over P}&\approx\beta{1-\log p\over p}&(\ref{effasym}b)\cr}$$
for current drive by Landau-damped and cyclotron-damped waves,
respectively.  [The factor of $1/v$ in Eqs.~(\ref{effloc}) is
replaced by unity in the limit $p\rightarrow\infty$.]
Equation~(\ref{effasym}a) (with $p$ replaced by $p_0$)
also applies for current drive by a narrow spectrum as given by
Eq.~(\ref{effnarrow}). In the limit of $p\rightarrow\infty$, the efficiency
of cyclotron-damped current drive vanishes, while for
current drive by Landau-damped waves
${J/P}\rightarrow \alpha$.  In order to determine this limiting efficiency,
either Eq.~(\ref{coeff}a) may be evaluated using the numerically
found value of $H_b(\Theta,Z)$ (see Table~\ref{eff-tab})
or else the equation may be
expanded as a series in $\Theta$ to give for $p\rightarrow\infty$
$${J\over P}\approx1+{5\over2}\Theta+H_b(0,Z)\Theta^{3/2}.\eqn(\eqlab{limit})$$
$H_b(0,Z)$ is tabulated in Table~\ref{coeff-tab}.

We now turn to the solution of Eq.~(\ref{spitz1}) in the
nonrelativistic limit $\Theta\rightarrow0$.  We shall still consider
only the limit $p\gg p_t$.  The limits here are nonuniform.
Equation~(\ref{spitz-norm}) was obtained by taking $p\gg p_t$ followed by
$\Theta\rightarrow0$.  Here we will take the limits in the opposite
order.  To do this, it is convenient to renormalize Eq.~(\ref{spitz1})
using $q$, $m$, $p_t$, and $\nu_t$ as the system of units.  In this case,
$J/P$ is normalized to $q/p_t\nu_t$.
Making this change of normalization and
taking the limit $\Theta\rightarrow0$ is equivalent to formally
replacing $\Theta$ by unity and substituting $v=p$, $\gamma=1$, and
$V_t^2=1$ in Eq.~(\ref{spitz-norm}) to give
$${1\over p^3}\biggl[ {\partial^2\chi_1\over \partial p^2}-
\biggr(p+{1\over p}\biggr)
{\partial\chi_1\over \partial p}\biggr]
-{1\over p^3}\biggl(1+Z-{1\over p^2}\biggr)\chi_1
+{H\over p^2}+p=0,
\eqn(\eqlab{nonrel})$$
where $H(Z)=H_a(0,Z)+H_b(0,Z)$ (this is tabulated in Table~\ref{coeff-tab}).
For $p\gg1$ (in this normalization this means
$p\gg p_t$), we may develop an asymptotic expression for $\chi_1$ as
a series in powers of $p$.  Balancing the terms
in Eq.~(\ref{nonrel}) from $O(p)$ (the leading order) to $O(p^{-4})$
gives
$$\chi_1= {p^4\over 5+Z}+{9p^2\over(5+Z)(3+Z)}
+{Hp\over 2+Z}+{9\over(5+Z)(3+Z)(1+Z)}+O(p^{-2}).$$
For localized excitation, Eq.~(\ref{effloc}) becomes
$$\eqalignno{{J\over P}&=
{4p^2\over 5+Z}+{18\over(5+Z)(3+Z)}
+{Hp^{-1}\over 2+Z}+O(p^{-4})&(\eqlab{effnonr}a)\cr
{J\over P}&={3p^2\over 5+Z}+{9\over(5+Z)(3+Z)}
-{9p^{-2}\over(5+Z)(3+Z)(1+Z)}+O(p^{-4})&(\ref{effnonr}b)\cr}$$
for Landau-damped waves and cyclotron-damped waves, respectively.  The
leading order terms here (those proportional to $p^2$) are exactly
those derived by Fisch and Boozer \cite{Fisch-Boozer}.

In order to compute the efficiencies for current drive
by a narrow spectrum of waves, it is necessary to carry out the
integrations in Eqs.~(\ref{effnarrow}) and~(\ref{effec}).  The
following asymptotic series is useful for this purpose:
$$\int_x^\infty \exp(-{\textstyle{1\over2}}y^2)y^{n+1}\,dy=
\exp(-{\textstyle{1\over2}}x^2)[x^{n}+nx^{n-2}+n(n-2)x^{n-4}+\cdots].$$
For $n$ even, the series terminates and is exact.
The efficiency for current drive by a narrow spectrum of
Landau-damped waves, Eq.~(\ref{effnarrow}) becomes
$${J\over P}={4v_p^2\over 5+Z}+{6(6+Z)\over(5+Z)(3+Z)}
+{Hv_p^{-1}\over 2+Z}+O(v_p^{-2}).\eqn(\eqlab{narrow-nonr})$$
For a narrow spectrum of cyclotron-damped waves, Eq.~(\ref{effec}) gives
$$\eqalignno{
{J\over P}&={3v_p^2\over 5+Z}+{3(9+2Z)\over(5+Z)(3+Z)}+O(v_p^{-2}),&
(\eqlab{narrow-ec}a)\cr
{J\over P}&={3v_p^2\over 5+Z}+{9(4+Z)\over(5+Z)(3+Z)}+O(v_p^{-2}),&
(\ref{narrow-ec}b)\cr}$$
for $l=1$ and $l=2$, respectively.  The effect of the integrations is to
change only the higher-order $O(v_p^0)$ corrections to the
efficiencies.  The leading order terms are the same as for the
localized excitation Eqs.~(\ref{effnonr}).
Equations~(\ref{narrow-nonr}) and~(\ref{narrow-ec}a) are
plotted in Fig.~\ref{nonrel-fig}.  These closely approximate the exact
results for $v_p>2v_t$

\section{HIGH ENERGY LIMIT OF COLLISION OPERATOR}\label{high-coll}

In the previous section, we derived finite temperature corrections to
the efficiency limit found in Ref.~\citenum{Fisch2}.  However, the
collision operator in the Landau form Eqs.~(\ref{coll}) was derived by
assuming that the background electrons are only weakly relativistic or
that $\Theta\ll1$.  We must check, therefore, that the finite $\Theta$
corrections to the Landau operator do not affect the formula for the
efficiency limit Eq.~(\ref{limit}).

The linearized collision operator Eq.~(\ref{lin-coll}) consists of three
collision terms.  Since in all practical cases the ions are
nonrelativistic, the ion term $C(f,f_i)$ needs no correction.  The
term $C(f_m,f)$ contributes to the integral term $I(\chi_1)$ in
Eq.~(\ref{spitz1}).  However, this resulted in a $O(\Theta^{3/2})$
contribution to efficiency limit Eq.~(\ref{limit}), so that corrections
to this term will be of still higher order.

Therefore, we need only consider collisions off a Maxwellian electron
background $C(f,f_m)$.  Furthermore, if $\Theta$ is small and if
$p\gg p_t$, we may take
$v'\ll v$ in the full collision kernel Eq.~(\ref{bb}) and approximate $\mat U$ by its Taylor expansion about
${\bf v}'=0$.  By retaining terms up to second order in ${\bf v}'$, we obtain
$$C(f,f_m)={\Gamma\over 2}{\partial\over\partial p_j}\biggl(
U_{jk}^{(0)}{\partial f\over\partial p_k}
+{\partial U_{jk}^{(0)}\over \partial v'_k} {v_t^2\over T}f
+{1\over2}{\partial^2 U_{jk}^{(0)}\over \partial v'_m\partial v'_m}
v_t^2{\partial f\over\partial p_k}\biggr)$$
where summation over repeated indices is implied
and the superscript $(0)$ is used to indicate that
$\mat U$ and its derivatives are evaluated at ${\bf v}'=0$.
Evaluating these coefficients gives
$$\eqalign{U_{jk}^{(0)}&={v^2\delta_{jk}-v_jv_k\over v^3},\cr
{\partial U_{jk}^{(0)}\over \partial v'_k}&=
{2v_j\over v^3},\cr
{1\over2}{\partial^2 U_{jk}^{(0)}\over \partial v'_m\partial v'_m}&=
{2v_jv_k\over v^5}-
(1-\beta^4){U_{jk}^{(0)}\over v^2}.\cr}$$
(This calculation was carried out using {\small MACSYMA} \cite{MACSYMA}.)
If we compare these with the equivalent expressions using $\mat U$
from Eq.~(\ref{coll}c), we find that only the term proportional to
$\beta^4$ is
new.  The high energy form of $C(f,f_m)$ is given by
Eq.~(\ref{isotrop}a) with $A(p)$ given by Eq.~(\ref{high}a),
$F(p)=(v/T)A(p)$, and
$$B(p)=\Gamma{1\over 2v}\biggl[1-{v_t^2\over v^2}
\biggl(1-{v^4\over c^4}\biggr)\biggr].$$
In other words, in the high-energy limit the electromagnetic correction
only changes the pitch-angle scattering term.
The new term has no effect on the asymptotic form for the
efficiencies Eqs.~(\ref{effasym}) because it is smaller by $\beta^4$
than another term in $B$, which had no effect.

Connor and Hastie \cite{Connor} also give an expression for collisions
of high-energy particles off a fixed background.  The corrections to
Eq.~(\ref{high}) that they obtain differ from ours.  This is possibly
because the background distribution that they treat is only
approximately Maxwellian.

\section{CONCLUSIONS}\label{conclusions}

We have considered the problem of current drive by fast waves in a
relativistic plasma.  Let us briefly review the approximations made.  The
major one is the reduction of the full collision operator to Landau
form.  We show in Sec.~\ref{collision} that this holds if the background
temperature is small, $T\ll mc^2$.  The corrections to the Landau
operator in the high energy limit are derived in Sec.~\ref{high-coll} and
are shown to be small.  The second important approximation is the
linearization of the electron-electron collision operator.  This is
accurate provided the rf strongly affects only electrons on the tail of
the distribution.  The subsequent analysis leading to the formula
for the current-drive efficiency Eq.~(\ref{effa}) is exact.  In order to
apply this formula, it is necessary to determine the rf-induced flux
$\bf S$ from Eq.~(\ref{flux}) and the Spitzer--H\"arm function $\chi$
from Eq.~(\ref{spitz}).

We considered two methods for computing $\bf S$: either to assume that
$f=f_m$ in Eq.~(\ref{flux}) (corresponding to linear damping) or to
solve the two-dimensional Fokker--Planck equation, Eq.~(\ref{2D}),
numerically.  The latter method may be necessary in the case of high
rf powers and wide spectra.  Note that the
efficiency can be accurately calculated even if the $\bf S$ is known
only approximately
since Eq.~(\ref{effa}), being an integral operator on $\bf S$,
is insensitive to small errors in $\bf S$.  Often, useful
information can be extracted from Eq.~(\ref{effa}) even with very
limited information about $\bf S$.  If the rf spectrum is
known, we can make some estimates (based on either numerical or
approximate analytical solutions to the Fokker--Planck equation) of
where in momentum space the flux is largest.  We can then use
Eq.~(\ref{effb}) to give the efficiency.

The Spitzer--H\"arm function $\chi$ can be determined by solving
Eq.~(\ref{spitz1}) numerically as in Sec.~\ref{numerical}.  Since this
equation is just a one-dimensional equation, there is little
difficulty in obtaining arbitrarily accurate results in this way.  This
method can be regarded as exact.  Alternatively, we found asymptotic
forms for $\chi$ in Sec.~\ref{asymptotics}.  From this we can write down
analytical expressions for the current-drive efficiency in various
cases as given in Eqs.~(\ref{effasym}), (\ref{limit}), (\ref{effnonr}),
(\ref{narrow-nonr}), and~(\ref{narrow-ec}).

The primary application of this work is, of course, to maintain a
steady-state toroidal current in a tokamak reactor.  The viability of
this scheme depends upon the amount of circulating power that is
required.  Thus, an accurate calculation of the current-drive
efficiency, as well as an assessment of the best possible efficiency,
are of crucial importance.

When applying these results to the study of steady-state current drive
in a tokamak, it is useful to convert the efficiency $J/P$ to $I/W$
where $I=AJ$ is the total current, $W=2\pi RAP$ is the total rf power,
$A$ is effective poloidal cross-sectional area, and $R$ is the tokamak
major radius.  This gives
$$\eqalignno{{I\over W}&={1\over 2\pi R}{J\over P}\cr
&=2.08{J/P\over q/mc\nu_c}{10^{20}\,{\rm m}^{-3}\over n}
{1\,{\rm m}\over R} {15\over\log\Lambda} \>{\rm A/W}\cr
&=40.7\times 10^{-3}{J/P\over q/p_t\nu_t}{10^{20}\,{\rm m}^{-3}\over n}
{T\over 10\,\hbox{keV}}
{1\,{\rm m}\over R} {15\over\log\Lambda} \>{\rm A/W}.\cr}$$
The last two equalities give the conversion from the normalized
efficiencies given in the figures and in Sec.~\ref{asymptotics} to
practical units.
Figures~\ref{local}, \ref{narrow}, and~\ref{nonrel-fig} contain scales
in these units.

The present work calculates the efficiency that can be expected from an
arbitrary wave-induced flux.  It is possible, therefore, to come to some
very general conclusions about the best possible efficiency that can be
obtained by driving currents with different waves.  In particular, there
is a limit, given by Eq.~(\ref{limit}),
to the efficiency of current drive with fast waves, such as lower-hybrid
waves, that interact through a Landau resonance with relativistic
electrons.  These waves are, perhaps, the most likely candidate for
current drive in a reactor.

The present calculations also apply to other types of current drive,
for example, relativistic electron beams \cite{REB}.  Here, the
efficiencies will be similar to those of Landau-damped waves.  Care must
be taken, however, in interpreting experiments on relativistic electron
beams because the assumption of a steady state is generally
inapplicable.

The equations developed here apply to other forms of rf current drive.
Some of these may be very efficient, more so than lower-hybrid
wave-induced fluxes.  For example, if low-phase-velocity waves interact
through a cyclotron resonance with fast electrons, the rf
flux may be nearly parallel to the constant energy contours, at the
same time that the collisionality of the resonant electrons is small.
This gives very high efficiency, but, in practice, these waves are much
more difficult to generate than are lower-hybrid waves.

Settling the question of the highest attainable current-drive efficiency
with fast waves should enable, we hope, tokamak reactor designers to
assess the practicality of using waves to drive steady-state currents.
There may, of course, be other effects that present difficulties, such
as the accessibility of the waves or nonlinear effects.  On the other
hand, there may be effects, such as the bootstrap current, which could
be helpful.

Finally, we hope that the form that we derived here for the
relativistic collision operator, which enabled us to solve for the
relativistic Spitzer--H\"arm function, will be of use in other numerical
problems dealing with collisions in hot plasmas.

\section*{ACKNOWLEDGMENTS}
This work was supported by the United States Department of Energy under
Contract DE--AC02--76--CHO--3073.

\begin{thetables}{99}
\newdimen\digitwidth\setbox0=\hbox{\rm0}\digitwidth=\wd0
{\catcode`?=\active\gdef\spdef{\offinterlineskip
 \catcode`?=\active\def?{\kern\digitwidth}}}
\tableitem{low-freq}  The coefficients for the efficiency for the three
types of current drive by low frequency waves.
$$\vcenter{\tabskip.5em\spdef
\halign{\strut\hfil$# $\hfil\tabskip2em&\hfil$# $\hfil
&\hfil$# $\hfil&\hfil$# $\hfil\tabskip.5em\cr\noalign{\hrule\vskip1.5pt\hrule}
Z    &    C_L     &    C_M     &    C_A     \cr\noalign{\hrule}
1    &    3.76    &    8.49    &    8.09    \cr
2    &    1.88    &    5.17    &    5.07    \cr
5    &    0.75    &    2.55    &    2.60    \cr
10   &    0.38    &    1.42    &    1.48    \cr
\noalign{\hrule\vskip1.5pt\hrule}}}$$
\tableitem{eff-tab}  Table of efficiencies ${J/P}$
for Landau-damped waves in the limit $v_p\rightarrow c$.
The efficiencies are normalized to $q/mc\nu_c$.
$$\vcenter{\tabskip.5em\spdef
\halign{\strut\hfil$# $\hfil
\tabskip2em&\hfil$# $\hfil&\hfil$# $\hfil&\hfil$# $\hfil&\hfil$# $\hfil
\tabskip.5em\cr\noalign{\hrule\vskip1.5pt\hrule}
\Theta &   Z=1   &   Z=2   &   Z=5   &   Z=10   \cr\noalign{\hrule}
 0.01  &   1.04  &   1.03  &   1.03  &   1.03   \cr
 0.02  &   1.09  &   1.07  &   1.06  &   1.06   \cr
 0.05  &   1.25  &   1.20  &   1.17  &   1.15   \cr
 0.1   &   1.55  &   1.44  &   1.34  &   1.30   \cr
 0.2   &   2.19  &   1.91  &   1.70  &   1.61   \cr
\noalign{\hrule\vskip1.5pt\hrule}}}$$
\tableitem{coeff-tab}  The coefficients $H_a(0,Z)$ and $H(Z)$.
$$\vcenter{\tabskip.5em\spdef
\halign{\strut\hfil$# $\hfil\tabskip2em&\hfil$# $\hfil
&\hfil$# $\hfil\tabskip.5em\cr\noalign{\hrule\vskip1.5pt\hrule}
  Z  &  H_a    &   H       \cr\noalign{\hrule}
  1  &  13.69  &   21.12   \cr
  2  &  ?9.13  &   13.51   \cr
  5  &  ?4.94  &   ?7.01   \cr
 10  &  ?2.88  &   ?4.01   \cr
\noalign{\hrule\vskip1.5pt\hrule}}}$$
\end{thetables}
\begin{thefigures}{99}
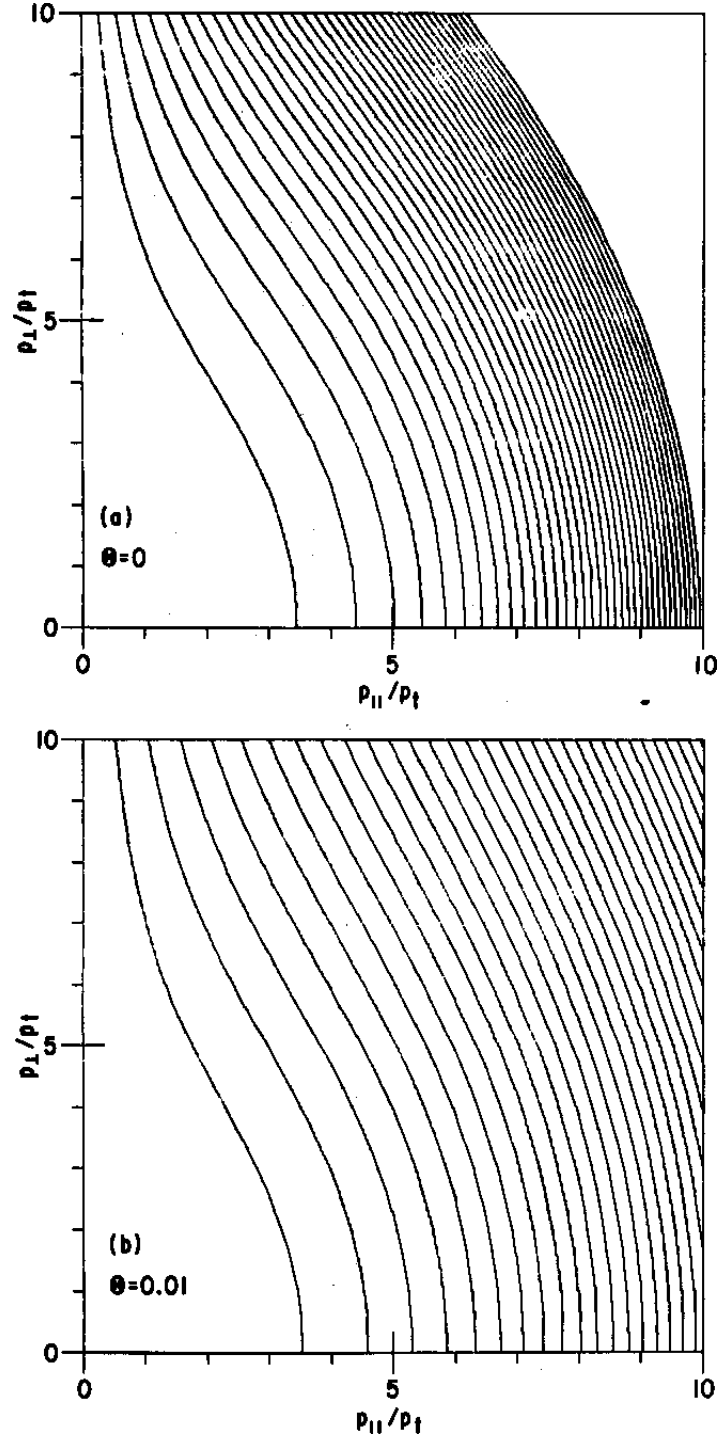
\figitem{contour}{3.7in} Contour plots of $\chi({\bf p})$
for $Z=1$ and (a) $\Theta=0$ and (b) $\Theta=0.01$.
The contour levels are evenly spaced with increments of
$50\,qp_t/m\nu_t$.  The higher
levels are on the right (i.e., $\partial\chi({\bf p})/\partial
p_\parallel>0$).
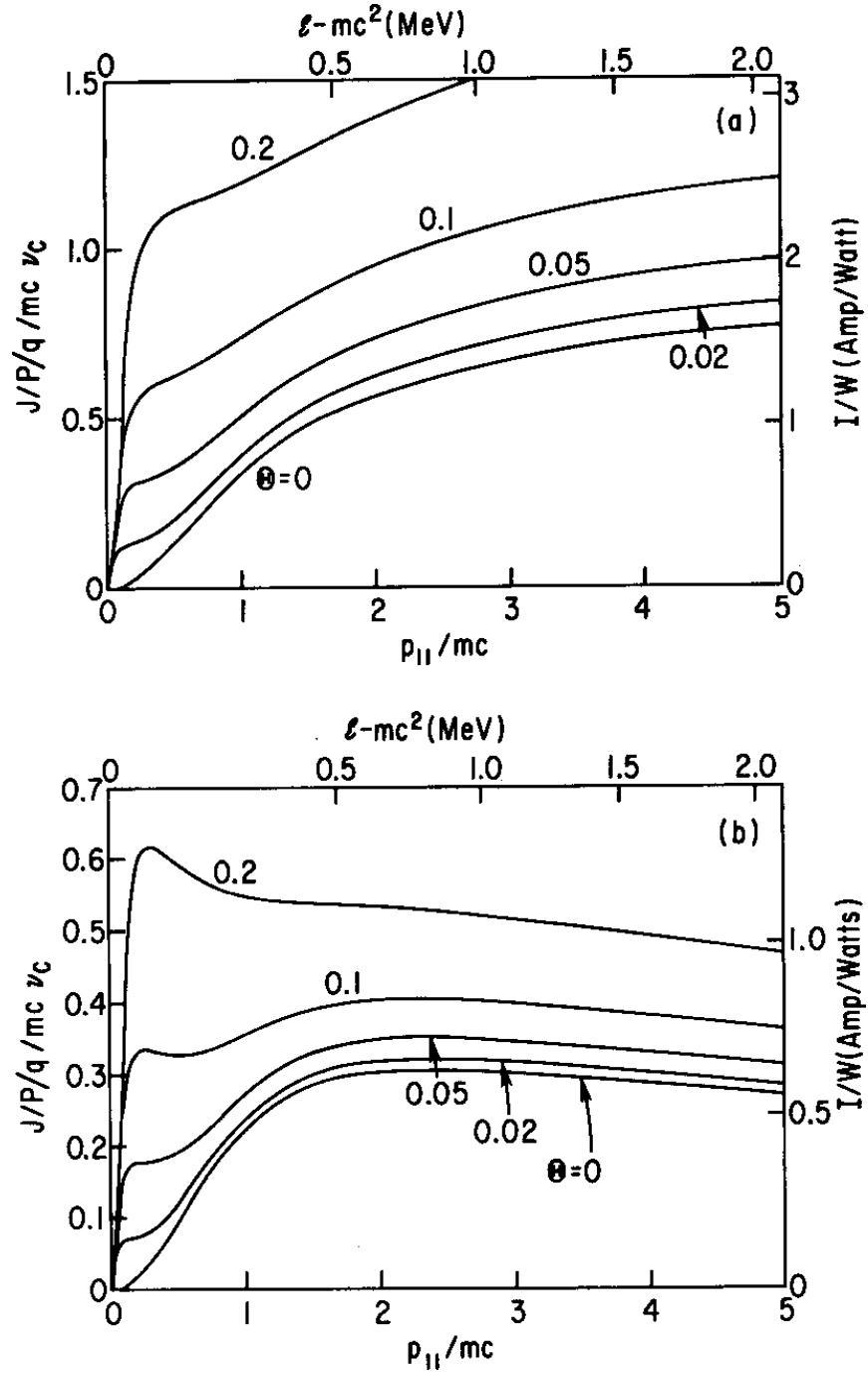
\figitem{local}{4.5in} Efficiencies for localized excitation
for (a) Landau-damped waves (parallel diffusion) Eq.~(\ref{effloc}a)
and (b) cyclotron-damped waves
(perpendicular diffusion) Eq.~(\ref{effloc}b).
The different curves show the efficiencies for various values of the
temperature $\Theta$ as indicated.  In all cases $Z=1$.  The top scale
gives the kinetic energy of the electrons.  The right scale gives the
efficiency for a plasma with $n=10^{20}\,{\rm m}^{-3}$,
$\log\Lambda=15$, and $R=1\,\rm m$.
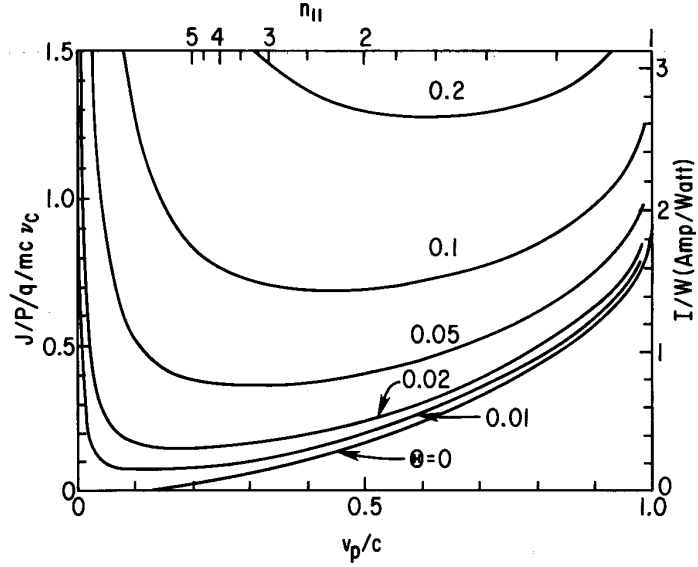
\figitem{narrow}{3.6in} Efficiencies for narrow Landau spectrum
Eq.~(\ref{effnarrow}) as a function of the phase velocity $v_p$.
The curves correspond to the various values of $\Theta$.
In all cases $Z=1$.  The top scale gives the parallel index
of refraction $n_\parallel=c/v_p$.  The right scale gives the
efficiency for the same conditions as in Fig.~\ref{local}.
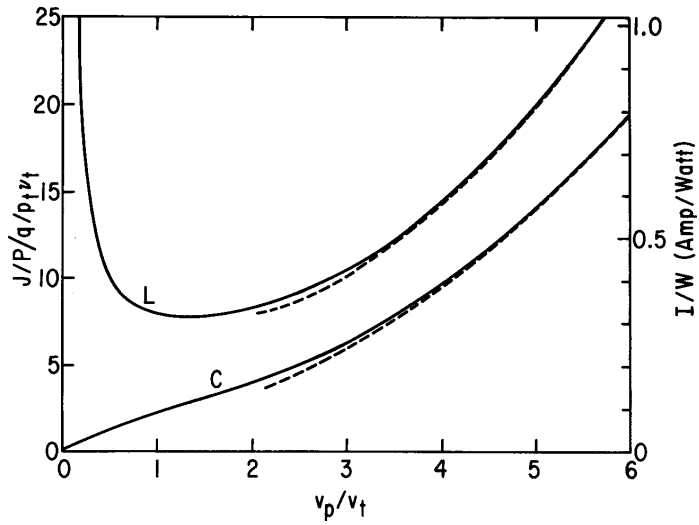
\figitem{nonrel-fig}{3.6in} Efficiencies for narrow spectra of Landau-damped
(L) waves and cyc\-lotron-damped (C) waves ($l=1$) for the
nonrelativistic case $\Theta\rightarrow0$ and $Z=1$.   Also shown as
dashed lines are
the asymptotic results Eqs.~(\ref{narrow-nonr}) and~(\ref{narrow-ec}a).
The right scale gives the
efficiency for a plasma with $n=10^{20}\,{\rm m}^{-3}$, $T=10\,\hbox{keV}$,
$\log\Lambda=15$, and $R=1\,\rm m$.
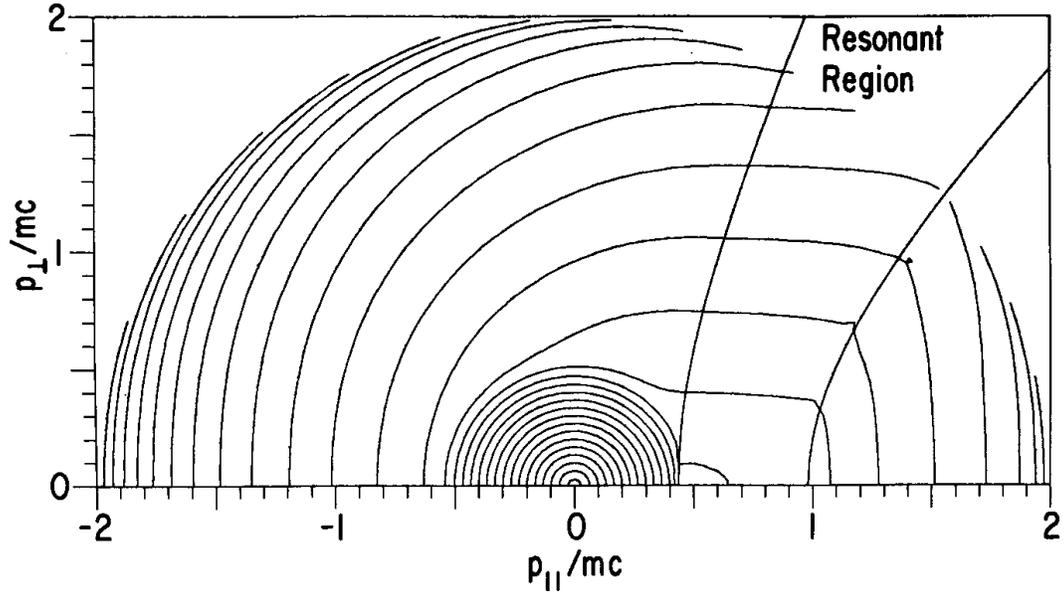
\figitem{2D plot}{5.5in} Contour plot of the steady-state distribution $f$
obtained by numerically integrating Eq.~(\ref{2D}).  Here $Z=1$,
$\Theta=0.01$, $v_1=0.4c$, $v_2=0.7c$. The resonant region is indicated.
The contour levels are chosen
so that for a Maxwellian they would be equally spaced with $\Delta
p=mc/30$.
\end{thefigures}
\end{document}